\documentstyle[twoside,fleqn,espcrc2,epsf]{article}
\title{The $\rho$ meson decay constant using a tadpole-improved action}
\author{Randy Lewis$^{\rm a}$ and 
        R. M. Woloshyn\address{TRIUMF, 4004 Wesbrook Mall, 
        Vancouver, British Columbia, Canada V6T 2A3}}
       
\begin{document}

\begin{abstract}
The $\rho$ meson decay constant and the associated renormalization factor
are computed in the quenched approximation on coarse lattices using a 
tadpole-improved action which is corrected at the classical level to 
${\cal O}(a^2)$.  The improvement is displayed by comparing to Wilson
action calculations.
\end{abstract}

\maketitle

\section{INTRODUCTION}

By improving a lattice action, one hopes that calculations at a fixed lattice
spacing will be closer to their continuum values than with the original action,
and similarly that calculations of a fixed accuracy can be performed on coarser
lattices than those required, for example, by the Wilson action.
For the present work, we consider an action which has classical errors 
beginning only at ${\cal O}(a^3)$, and which is also 
tadpole-improved\cite{LM}.  This action has been used to compute
the spectrum of light hadron masses\cite{FW,LL}, and it was found that hadron
mass ratios are considerably improved compared to Wilson action results at
comparable lattice spacings.  In this work we address the question of whether
matrix elements are similarly improved by considering the vector current
renormalization factor and vector meson decay constant.

\section{IMPROVED ACTION}

We use an improved gauge field action which involves a sum over 1$\times$2
plaquettes as well as 1$\times$1 plaquettes\cite{LW},
and an improved fermion action which contains next-nearest-neighbour 
interactions\cite{fermionaction},
\begin{eqnarray}
&& \!\!\!\!\!\!\!\!\!\!\!\!\! 
S_F(\bar{\psi},\psi;U)  =  -\sum_{x}\bar{\psi}(x)\psi(x) \nonumber \\
&& ~~~~~\,    +\frac{4}{3}\kappa\sum_{x,\mu}\left[
\bar{\psi}(x)(1-\gamma_{\mu})U_{\mu}(x)\psi(x+\mu)\right. \nonumber \\
&& ~~~~~~~~
\phantom{\frac{4}{3}\kappa\sum_{x,\mu}}
       \left. +\bar{\psi}(x+\mu)(1+\gamma_{\mu})
U_{\mu}^{\dagger}(x)\psi(x)\right] \nonumber \\
&& \!\!\!\!\!\!\!\!\!\!\!\!\!    -\frac{\kappa}{6U_0}\sum_{x,\mu}
\left[\bar{\psi}(x)(2-\gamma_{\mu})
U_{\mu}(x)U_{\mu}(x+\mu)\psi(x+2\mu)\right. \!\!\!\!\!\!\!\!\!\!\nonumber \\
&& \!\!\!\!\!\!\!\!\!\!\!\!\!  \left. +\bar{\psi}(x+2\mu)(2+\gamma_{\mu})
   U_{x}^{\dagger}(x+\mu)U_{\mu}^{\dagger}(x)\psi(x)\right]~.
\end{eqnarray}
The tadpole factor\cite{LM} is 
$U_0 = \langle\frac13\mbox{Re}\mbox{Tr}U_{pl}\rangle^{1/4}$.

The presence of three timesteps in a single term of the action introduces
artifacts\cite{artifacts}, such as oscillations in meson mass functions near 
the source.  However on our lattices, the plateau itself is not altered.

\section{METHOD}

In the continuum, the $\rho$ meson decay constant $f_\rho$ is defined as
follows:
\begin{equation}
   \langle 0|V_\mu|\rho\rangle_{\rm cont} =
   f_\rho^{-1}m_\rho^2\epsilon_\mu~.
\end{equation}
On a lattice, matrix elements of the vector current get
renormalized by a factor $Z_V$,
\begin{equation}
   \langle f|V_\mu|i\rangle_{\rm cont} =
               Z_V(g^2)\langle f|V_\mu|i\rangle + \ldots~,
\end{equation}
where the dots represent finite lattice spacing effects.
For the improved action, the ${\cal O}(a,a^2)$ terms vanish identically and 
terms proportional to the coupling are kept small by
the tadpole factor.  By considering the ratio of the local vector current
$V_\mu^L(x) = \overline{\psi}(x)\gamma_\mu\psi(x)$
to the conserved vector current $V_\mu^C(x)$,
we can compute the renormalization factor,
\begin{equation}\label{ratio}
   Z_V(g^2) =
   {\langle f|V_\mu^C|i\rangle}\,/\,{\langle f|V_\mu^L|i\rangle}\, +\, \ldots~.
\end{equation}
We will consider a variety of choices for the initial and final states in this
ratio, and the differences in the resulting values for $Z_V$ will reflect the 
finite lattice spacing effects.

\section{CALCULATION}

We have performed calculations at two values of $\beta$ with the improved 
action which correspond to lattice spacings of 0.4fm and 0.27fm as derived from
the string tension.  For comparison, we have also used the Wilson action to 
calculate at the same lattice spacings.  For the gauge fields we used
pseudo-heatbath updating with periodic boundary conditions; the first 4000
sweeps were discarded and then 250 sweeps (200 sweeps) were discarded between 
each pair of improved (Wilson) configurations that was kept.

A stabilized biconjugate gradient algorithm was used for the fermion matrix 
inversion at three choices of the hopping parameter.  Due to the modest
number of lattice sites, Dirichlet boundary
conditions were used for fermions propagating in the time direction but 
periodic boundaries were used in all spacial directions.  The source was
placed two timesteps away from the boundary.

\begin{table}[hbt]
\setlength{\tabcolsep}{.33pc}
\newlength{\digitwidth} \settowidth{\digitwidth}{\rm 0}
\caption{Some details of the simulations.  $N_U$ is the number of gauge field 
         configurations and $a_{st}$ is the lattice spacing derived from the 
         string tension.}
\label{tab:details}
\begin{tabular*}{75mm}{@{\extracolsep{\fill}}ccccc} 
\hline
\rule[-2mm]{0mm}{6mm} Lattice & $N_U$ & $\beta$ & $a_{st}$[fm] & $\kappa$ \\ 
\hline
\multicolumn{5}{c}{\rule[-2mm]{0mm}{8mm}Improved Action} \\
 $6^3$${\times}12$ & 100 & 6.25 & 0.4  & 0.162, 0.168, 0.174 \\
 $8^3$${\times}14$ &  50 & 6.8  & 0.27 & 0.150, 0.154, 0.158 \\
\multicolumn{5}{c}{\rule[-2mm]{0mm}{8mm}Wilson Action} \\
 $6^3$${\times}12$ &  75 & 4.5  & 0.4  & 0.189, 0.201, 0.213 \\
 $8^3$${\times}14$ &  30 & 5.5  & 0.27 & 0.164, 0.172, 0.180 \\ 
\hline
\end{tabular*}
\end{table}

\section{RESULTS}

\begin{figure}[thb]
\epsfxsize=200pt \epsfbox[60 489 528 732]{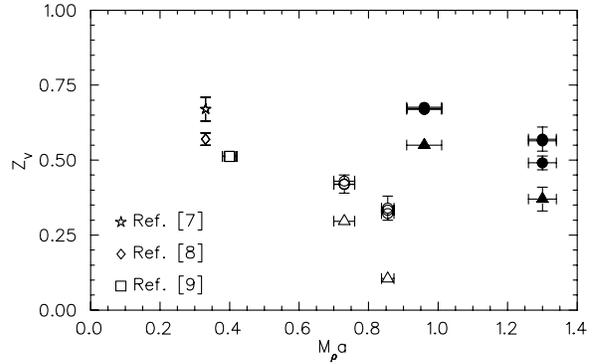}
\vspace{9pt}
\caption{The renormalization constant in the chiral limit.  Solid (open) 
         symbols are from the improved (Wilson) action.
         The lowest lying point at each 
         $M_\rho a$ is from the decay amplitude.}
\label{fig:totZ}
\end{figure}

\begin{figure}[bht]
\epsfxsize=200pt \epsfbox[60 489 528 732]{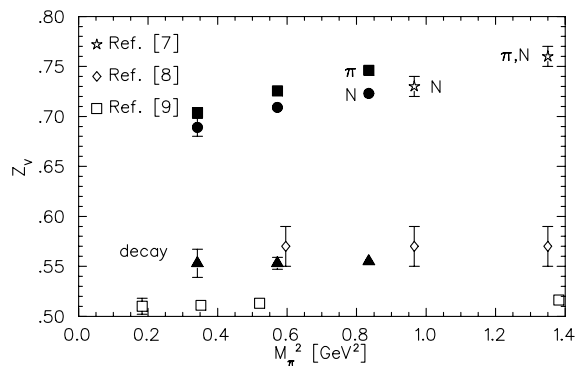}
\vspace{9pt}
\caption{The renormalization constant as a function of $M_\pi^2$.  Solid 
         symbols are from the improved action at $\beta = 6.8$;  
         open stars and diamonds are Wilson at $\beta = 6.0$; open squares
         are Wilson at $\beta = 5.85$.}
\label{fig:sinZ}
\end{figure}

Figure~\ref{fig:totZ} shows our determination of $Z_V$
from four separate choices of the states in Eq.~(\ref{ratio}), 
\[
   <\pi|V|\pi>, <\rho|V|\rho>, <N|V|N>, <0|V|\rho>.
\]
The extrapolation to the chiral limit was done linearly in $m_\pi^2$.
The results of Wilson calculations by other groups are also shown.  Notice
that for both actions, $Z_V$ at fixed $M_\rho a$ is approximately independent 
of the amplitude 
used, except for the decay amplitude which predicts a smaller value for $Z_V$.
For fixed lattice spacing, the improved action does give 
improved values of $Z_V$; in fact the improved
action matches the Wilson predictions for $Z_V$ when the lattice spacing is
larger by a factor of about three.  
Figure~\ref{fig:sinZ} shows that the dependence on $M_\pi^2$ is also quite 
similar for the two actions.

\begin{figure}[htb]
\epsfxsize=200pt \epsfbox[60 489 528 732]{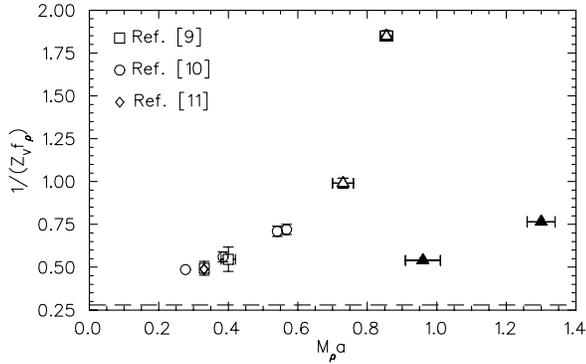}
\vspace{9pt}
\caption{The lattice decay constant.  The dashed line is the experimental
         value; solid symbols are from the improved 
         action; open symbols are Wilson calculations.  
         All Wilson quarks are normalized by $\sqrt{2\kappa}$.}
\label{fig:fVnoZ}
\end{figure}
Figure~\ref{fig:fVnoZ} shows the {\it lattice\/} decay constant
(i.e. without $Z_V$).
Again, we find that the improved action does give improved results
when compared to Wilson calculations at the same lattice spacing, and that
the improved prediction is similar to the Wilson prediction at 1/3 of the
lattice spacing.

\begin{figure}[htb]
\epsfxsize=200pt \epsfbox[60 489 528 732]{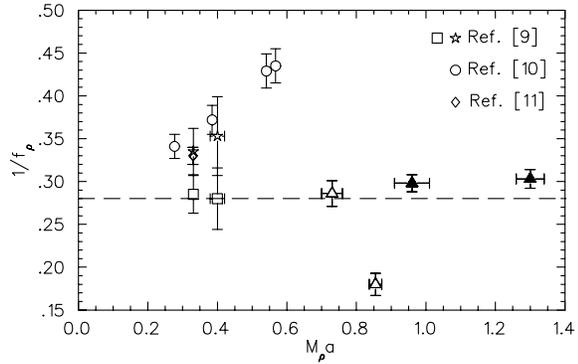}
\vspace{9pt}
\caption{The renormalized decay constant.  The dashed line is the experimental
         value; solid symbols are from the improved 
         action; open triangles and open squares are from the Wilson action. 
         All other symbols are Wilson calculations with $Z_{\rm pert}$.}
\label{fig:fV}
\end{figure}

Our best determination of the decay constant comes from the product of $Z_V$ 
and
$1/(Z_Vf_\rho)$, and this is shown in Figure~\ref{fig:fV}.  The cluster of data
points which lie above the experimental result rely on tadpole-improved
perturbation theory to estimate $Z_V$.  Notice that the completely 
nonperturbative determinations of $1/f_\rho$ are quite flat over a substantial
range of lattice spacing, and that for both the Wilson and improved actions
the result is near experiment.

In summary, we have found that the ${\cal O}(a^2)$ classical and tadpole 
improvements of our action manifest themselves in vector current matrix 
elements
in essentially the same way as in masses\cite{FW}.  Improved action results
for $1/f_\rho$ and $Z_V$ are consistent with Wilson action computations done
at a lattice spacing about 1/3 the size.

\section*{Acknowledgements}

This work was supported in part by the Natural Sciences and Engineering
Research Council of Canada.

\end{document}